\documentclass[showpacs,10pt,twocolumn,prb]{revtex4-1}

\usepackage{amsmath}
\usepackage{amssymb}
\usepackage{graphicx}
\usepackage{amssymb}
\usepackage{graphics}
\usepackage{epsfig}
\usepackage{CJK}
\usepackage{color}

\begin{document}

\begin{CJK*}{GBK}{}
\title{Large magnetoresistance in type-$\rm{\uppercase\expandafter{\romannumeral2}}$ Weyl semimetal WP$_2$}
\author{Aifeng Wang,$^{1}$ D. Graf,$^{2}$ Yu Liu,$^{1}$ Qianheng Du,$^{1,3}$ Jiabao Zheng,$^{4,5}$ Hechang Lei,$^{6}$ and C. Petrovic$^{1,3}$}
\affiliation{$^{1}$Condensed Matter Physics and Materials Science Department, Brookhaven National Laboratory, Upton, New York 11973, USA\\
$^{2}$National High Magnetic Field Laboratory, Florida State University, Tallahassee, Florida 32306-4005, USA\\
$^{3}$2Department of Materials Science and Chemical Engineering, Stony Brook University, Stony Brook, New York 11790, USA\\
$^{4}$Department of Electrical Engineering, Columbia University, New York, New York 10027,USA\\
$^{5}$Department of Electrical Engineering and Computer Science, Massachusetts Institute of Technology, Cambridge, Massachusetts
02139, USA\\
$^{6}$Department of Physics and Beijing Key Laboratory of Opto-electronic Functional Materials and Micro-nano Devices, Renmin University of
China, Beijing 100872, People's Republic of China}

\date{\today}

\begin{abstract}
We report magnetotransport study on type-II Weyl semimetal WP$_2$ single crystals. Magnetoresistance (MR) exhibits a nonsaturating $H^{n}$ field dependence (14,300\% at 2 K and 9 T) whereas systematic violation of Kohler's rule was observed. Quantum oscillations reveal a complex multiband electronic structure. The cyclotron effective mass close to the mass of free electron m$_e$ was observed in quantum oscillations along $b$-axis, while reduced effective mass of about 0.5$m_e$ was observed in $a$-axis quantum oscillations, suggesting Fermi surface anisotropy. Temperature dependence of the resistivity shows a large upturn that cannot be explained by the multi-band magnetoresistance of conventional metals. Even though crystal structure of WP$_{2}$ is not layered as in transition metal dichalcogenides, quantum oscillations suggest partial two-dimensional character.

\end{abstract}
\maketitle
\end{CJK*}

\section{INTRODUCTION}

Weyl fermions originate from Dirac states by breaking either time-reversal symmetry or space-inversion symmetry.\cite{Dirac,WanX,Balents,Soluyanov,VafekO,BansilA,HasanM,JiaS,BurkovA,LiuJ,ZhengH,DaiX} In addition to  type-$\rm{\uppercase\expandafter{\romannumeral1}}$\cite{Balents,BansilA,LiuJ} Weyl points with a close point-like Fermi surface, novel type-$\rm{\uppercase\expandafter{\romannumeral2}}$\cite{Soluyanov,ZhengH} Weyl fermions appear at the boundaries between electron and hole pockets violating Lorentz invariance. This results in an open Fermi surface and anisotropic chiral anomaly. Most of the reported type-$\rm{\uppercase\expandafter{\romannumeral2}}$ Weyl semimetals exhibit characteristic extraodinary magnetoresistance (XMR), observed for example in materials that crystallize in C2/m space group of OsGe$_2$-type structure, such as NbSb$_2$,  TaAs$_2$, and NbAs$_2$.\cite{WangKF,YuanZ}

\begin{figure}
\centerline{\includegraphics[scale=0.4]{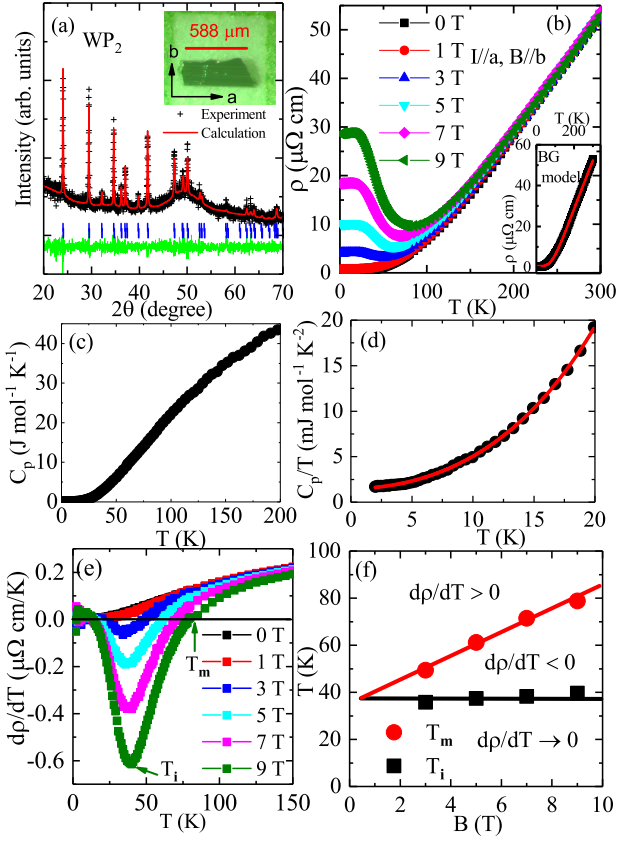}}
\caption{(Color online). (a) Powder X-ray diffraction pattern of WP$_2$, inset shows the photograph of as grown single crystal.  (b) Temperature-dependent transversal resistivity for WP$_2$ in different magnetic fields. Inset shows $\rho$(0T) fitted with Bloch-Gr\"{u}neisen model. Heat capacity of WP$_{2}$ (c,d). The($\partial\rho/\partial T$) plotted as a function of temperature (e); minima and sign change points in the $\partial\rho/\partial T$ are defined as $T_m$ and $T_i$, respectively. Magnetic field dependence of $T_m$ and $Ti$ (f).}
\label{magnetism}
\end{figure}

WP$_2$ crystallizes the nonsymmorphic space group Cmc2$_1$(36), which favors new topological phases with Dirac states.\cite{Rundquist,YoungS,BzdusekT,ZhaoYX} Indeed, WP$_2$ was recently predicted to be a type-$\rm{\uppercase\expandafter{\romannumeral2}}$ Weyl semimetal with four pairs of type-$\rm{\uppercase\expandafter{\romannumeral2}}$ Weyl points and long topological Fermi arcs.\cite{Autes} In this work, we have successfully grown single crystals of WP$_2$, and have investigated its magnetotransport properties. Magnetoresistance (MR = [$\rho$(B) - $\rho$(0)]/$\rho$(0)$\times$100\%) exhibits a nonsaturating $\sim$$H^{1.8}$ field dependence up  to 14,300\% at 2 K and 9 T. WP$_2$ shows a systematic violation of Kohler's rule and large magnetic field-induced nonmetallic resistivity that cannot be explained by orbital magnetoresistance of multiband metals. Fermi surface studies of of WP$_2$ by quantum oscillations reveal complex electronic structure. The $T - H$ phase diagram is consistent with the universal phase diagram for extraordinary magnetoresistance (XMR) materials,\cite{TaftiPNAS} however the Weyl points in WP$_{2}$ have little influence on the effective mass and mobility of the Fermi surfaces detected by quantum oscillations.

\section{EXPERIMENTAL DETAILS}

Single crystals of WP$_2$ were grown by chemical vapor transport method using I$_2$ as transport agent. First, WP$_2$ polycrystals were obtained by annealing stoichiometric W and P powder at 500 $^{\circ}$C for 24 h, and then 750 $^{\circ}$C for 48 h. WP$_2$ polycrystals were mixed with I$_2$ (15 mg/ml), and then sealed in an evacuated quartz tube. Single crystals were grown in a temperature gradient 1180 $^{\circ}$C (source) to 1050 $^{\circ}$C (sink) for a month. Shiny needle-like single crystal with typical size 1.5 mm $\times$ 0.05 mm $\times$ 0.05 mm were obtained. X-ray diffraction (XRD) measurements were performed using a Rigaku Miniflex powder diffractometer. The element analysis was performed using an energy-dispersive x-ray spectroscopy (EDX) in a JEOL LSM 6500 scanning electron microscope. Heat capacity and magnetotransport measurement up to 9 T were conducted in a Quantum Design PPMS-9. Magnetotransport measurement at high magnetic fields up to 18 T were conducted at the National High Magnetic Field Labortory (NHMFL) in Tallahassee. Resistivity was measured using a standard four-probe configuration. Hall resistivity was measured by the four-terminal technique by switching the polarity of the magnetic field to eliminate the contribution of $\rho_{xx}$ due to the misalignment of the voltage contacts.

\section{RESULTS AND DISCUSSIONS}

\begin{figure}
\centerline{\includegraphics[scale=0.35]{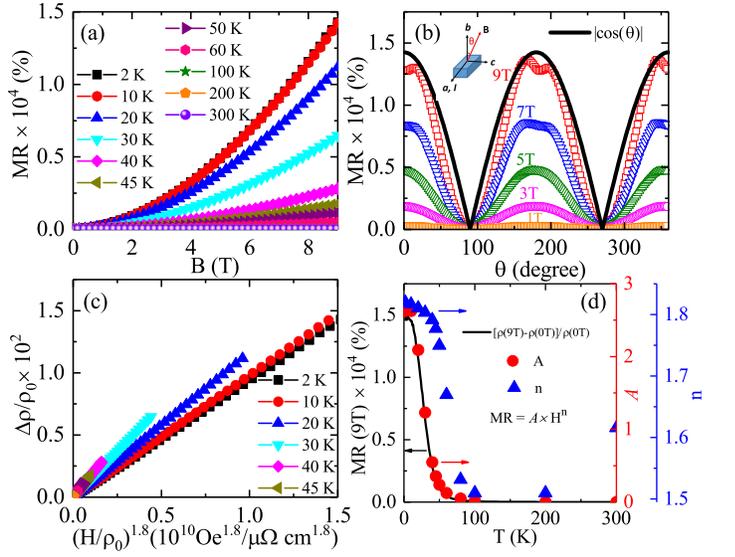}}
\caption{(Color online). Magnetic field dependence of MR at various temperatures (a) and magnetoresistance as a function of the titled angle from the applied field (b). Dashed line is the fitting using the 2D model. Kohler plot at different temperatures using $\Delta\rho$/$\rho(0)$ = $F[H/\rho(0)]$ $\simeq$ $H^{1.8}$ (c). MR versus temperature for WP$_2$ at 9 T (left) and temperature dependence of $A$ and $n$ (MR = $A \times H^n $) (right) (d).}
\label{magnetism}
\end{figure}

Figure 1(a) shows the  powder X-ray diffraction pattern of WP$_2$. The data can be fitted quite well by the space group Cmc2$_1$ with lattice parameter $a$ = 0.3164(2) nm, $b$ = 1.184(2) nm, and $c$ = 0.4980(2) nm, consistent with previous report.\cite{Rundquist} The average atomic ratio determined by the EDX is W: P = 1: 2. Figure 1(b) shows the temperature dependence of resistivity in WP$_2$ in different fields. The residual resistivity ratio (RRR) in the absence of magnetic field is $\rho(300K)$/$\rho(2K)$ = 279 with $\rho(2K)$ = 0.19 $\mu\Omega$cm, indicating small defect scattering in the crystal. The metallic resistivity in $H$ = 0 can be described by the Bloch-Gr\"{u}neisen (BG) model:\cite{Ziman}
\[
\rho (T)=\rho _0 + C(\frac{T}{\Theta_D})^5\int_{0}^{\frac{\Theta_D}{T}}\frac{x^5}{(e^x-1)(1-e^{-x})}dx
\]
where $\rho_0$ is the residual resistivity and $\Theta_D$ is the Debye temperature, indicating phonon scattering in the whole temperature range. The fitting results in $\Theta_D$ = 546(5) K. As shown in Fig. 1(b), with the application of the magnetic field, the $\rho$(T) shows a large upturn which saturates below $\sim$ 15 K. Similar resistivity behaviour was observed recently in XMR materials such as WTe$_2$, TaAs, LaSb, PtSn$_4$, ZrSiSe, Ta$_3$S$_2$, but also in graphite, bismuth, and MgB$_2$.\cite{Tafti,Hechang,Cava,HuangX,MunE, Hosen, ChenD,Soule,ZhuZ,BudkoS} Heat capacity of WP$_{2}$, shown in Fig. 1(c), is best described at low temperatures with a $C(T)/T=\gamma+\beta
T^{2}+\delta T^{4}$ where $\gamma$ = 1.50(5) mJ mol$^{-1}$ K$^{-1}$, $\beta$ = 0.0324(5) mJ mol$^{-1}$ K$^{-2}$ and $\delta$ = 3.0(2) $\times$ 10$^{-5}$ mJ mol$^{-1}$ K$^{-4}$ [Fig. 1(d)].\cite{Cezairliyan} The Debye temperature $\Theta_{D}=564(\pm11)$ K is obtained from $\Theta_{D}=(12\pi^{4}NR/5\beta)^{1/3}$ where N is the atomic number in the chemical formula and R is the gas constant.

Based on the field dependence of $\partial\rho/\partial T$ [Fig. 1(e)] we plot the temperature-field ($T-H$) phase diagram in Fig. 1(f), where $T_m$ and $T_i$ are taken as the sign change point and minimum in $\partial\rho/\partial T$. The phase diagram is consistent with extreme magnetoresistance materials.\cite{TaftiPNAS}

\begin{figure}
\centerline{\includegraphics[scale=0.22]{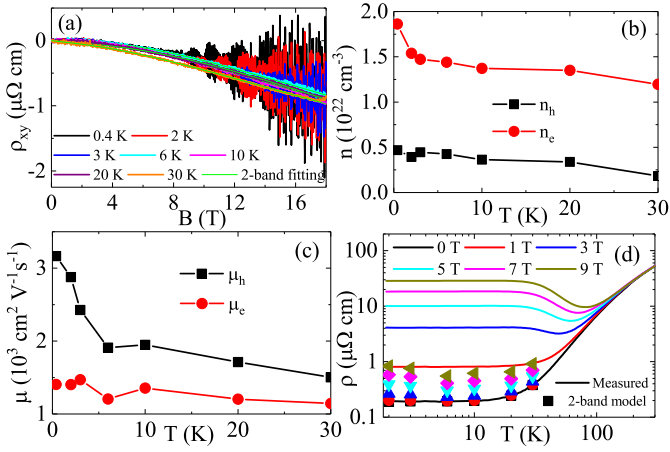}}
\caption{(Color online). (a) The magnetic field dependence of Hall resistivity $\rho_{xy}$ at various temperatures. The temperature dependence of carrier densities (b) and mobilities (c) of electrons and holes obtained by fitting using two band model. (d) Measured temperature dependence of $\rho$ (T, B) at various field (solid lines), and $\rho$ (T, B) calculated using the temperature dependence of $n_{e,h}$ and $\mu_{e,h}$ derived from the two-band model (scatter symbols).}
\label{Hall}
\end{figure}

Figure 2(a) shows the field dependence of transverse magnetoresistance MR = [$\rho$(B) - $\rho$(0)]/$\rho$(0)$\times$100\% at various temperatures with B//$b$. MR exhibits a non-saturating $\sim$$H^{1.8}$ dependence and reaches 14,300\% at 2 K and 9 T. It decreases quickly with increasing temperature, and becomes negligibly small above 100 K. The MR of WP$_2$ is about one order of magnitude smaller than in WTe$_2$ and LaSb, but comparable even larger than those in LaBi, Ta$_3$S$_2$, and TaIrTe$_4$.\cite{Cava,Tafti,Hechang, ChenD, Khim} The MR of solids only responds to the extremal cross section of the Fermi surface along the field direction. Thus, angular dependence of MR in a material with a 2D Fermi surface should be proportional to $\mid$cos($\theta$)$\mid$.\cite{Shoeneberg} As shown in Fig. 2(b), MR in different magnetic fields show two fold symmetry which can be fitted well with $\mid$cos($\theta$)$\mid$, indicating a 2D Fermi surface.

Field induced nonmetallic resistivity and XMR in topological materials has been discussed lately in the literature; however its origin was ascribed to different mechanisms. In WTe$_2$ it was suggested that the band structure of a compensated semimetal with perfect electron-hole symmetry is important whereas in LaBi and LaSb the proposed mechanism involves a combination of compensated electron-hole pockets and particular orbital texture on the electron pocket.\cite{TaftiPNAS,WangL,NiuXH,WuY,NayakJ} There are also evidences for exotic and multiple surface Dirac states in other materials. MR in PtSn$_4$ has been considered within the model of orbital MR in the long mean-free path metals with single dominant scattering time.\cite{MunE} However, Dirac node arcs are also found in PtSn$_4$ and ZrSiSe(Te), suggesting that XMR could be connected with suppressed backscattering of Dirac states.\cite{WuY2,MaoZQ}

\begin{figure}
\centerline{\includegraphics[scale=0.3]{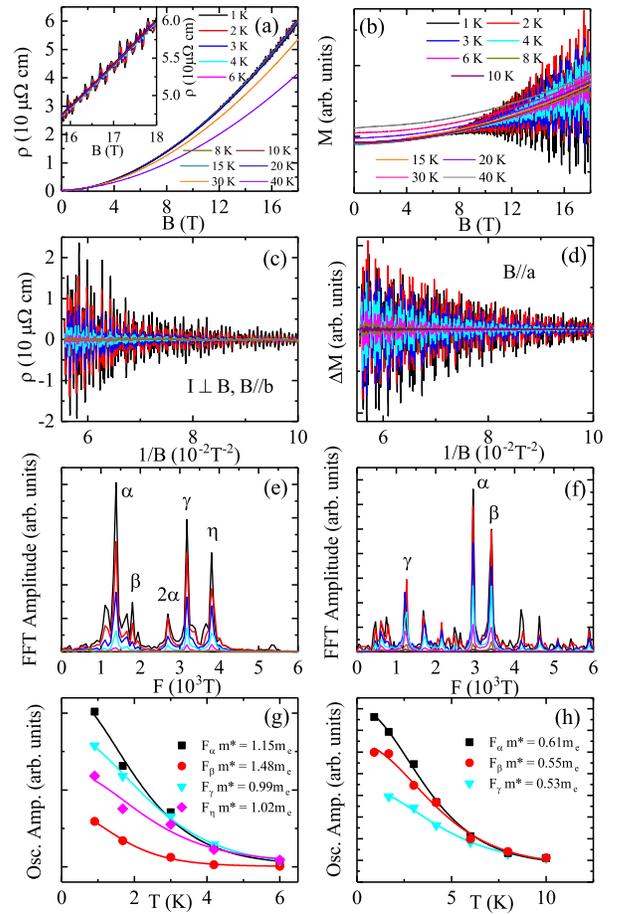}}
\caption{(Color online). Transverse resistivity vs field ($B//c$) at different temperatures (a); inset shows the enlarged part in high magnetic field. Magnetization ($B//a$) measured up to 18 T at various temperatures (b). SdH (c), dHvA (d) oscillatory components obtained by smooth background subtraction. FFT spectra for SdH (e)and dHvA (f), respectively. Temperature dependence of oscillating amplitude at different frequencies for SdH(g) and dHvA(h), respectively, Solid lines are fitted using Lifshitz-Kosevich formula.}
\label{oscillation}
\end{figure}

Semiclassical transport theory based on the Boltzmann equation predicts Kohler's rule $\Delta\rho/\rho$(0) = $F[H/\rho(0)]$ to hold if there is a single type of charge carrier and scattering time in a metal.\cite{Pippard} Violation of Kohler's rule is common in XMR materials, such as LaBi, TaAs, TaAs$_2$, NbAs$_2$, NbSb$_2$.\cite{Hechang,ZhangCL,YuanZ,WangKF}  As shown in Fig.2(c), MR in WP$_2$ systematically deviates from Kohler's rule above 10 K. The field dependence of MR at different temperature can be well fitted with MR = $A$ $\times$ $H^n$, $n$ is 1.8 at 2 K. The exponent $n$ systematically decreases to 1.5 with increasing temperature, as shown in Fig.2(d). The temperature evolutions of $A$ and $n$ are similar to temperature dependence of MR(9T).

A common violation of Kohler's rule in metals is multi-band electronic transport i.e. existence of multiple scattering times. WP$_2$ was predicted to be a type-$\rm{\uppercase\expandafter{\romannumeral2}}$ Weyl semimetal; therefore there is a possibility that high mobility bands with Dirac states contribute to small residual resistivity in zero magnetic field and strong magnetoresistance. In the two-band model MR is:\cite{SmithR}

\[\begin{array}{l}
MR = \frac{n_e\mu_en_h\mu_h(\mu_e + \mu_h)^2(\mu_0H)^2}{{{{({\mu _e}{n_h} + {\mu _h}{n_e})}^2} + {{({\mu _h}{\mu _e})}^2}{{({\mu _0}H)}^2}{{({n_h} - {n_e})}^2}}}
\end{array}\]

For the compensated semimetal, where $n_e$ $\simeq$ $n_h$, we obtain MR = $\mu_e\mu_h$$(\mu_0H)^2$. Since MR = $A$ $\times$ $H^n$ ($n$ = 1.5-1.8) [Fig. 2(c)] compensated two band model cannot completely explain the MR. Next, we discuss electronic transport in WP$_{2}$ from the perspective of two-band model when $n_e$ $\neq$ $n_h$.

Figure 3(a) shows the field dependence of Hall resistivity $\rho_{xy}$ at different temperatures. Clear quantum oscillation were observed at low temperatures. The frequencies and effective mass derived from $\rho_{xy}$ are consistent with that of $\rho_{xx}$, and we discuss this below. The $\rho_{xy}$ can be fitted by two band model:

\[\begin{array}{l}
\frac{{{\rho _{xy}}}}{{{\mu _0}H}} = \mathop R\nolimits_H {\rm{ }}
 = \frac{1}{e}\frac{{(\mu _h^2{n_h} - \mu _e^2{n_e}) + {{({\mu _h}{\mu _e})}^2}{{({\mu _0}H)}^2}({n_h} - {n_e})}}{{{{({\mu _e}{n_h} + {\mu _h}{n_e})}^2} + {{({\mu _h}{\mu _e})}^2}{{({\mu _0}H)}^2}{{({n_h} - {n_e})}^2}}}
\end{array}\]

Where $n_{e}$ ($n_h$) and $\mu_{e}$ ($\mu_h$) denote the carrier concentrations and mobilities of electron and hole, respectively. We find that $\rho_{xy}$ can be well fitted by the two-band model. The obtained $n_{e}$ ($n_h$) and $\mu_{e}$ ($\mu_h$) by fitting are shown in Fig. 3(b) and 3(c), respectively. WP$_2$ shows relatively high carrier density and the dominant carrier is electron-like. The mobility is similar to other Dirac materials such as $A$MnBi$_{2}$ ($A$ = Sr,Ca), LaBi and doped WTe$_{2}$.\cite{KefengCa,KefengSr,Hechang,KrienerM} The large carrier density indicates that WP$_2$ is different from a compensated semimetal. We calculate the $\rho \; (T, B)$ using carrier concentrations $n_{e}$ ($n_h$) and mobilities $\mu_{e}$ ($\mu_h$) obtained by two band model fitting of $\rho_{xy}$, as shown in in Fig.3(d). A similar temperature dependence between calculated and measured MR argues in favor of the validity of multicarrier electronic transport [Fig. 3(d)]. However, in contrast to LaBi, the $\rho \; (T, B)$ from two-band orbital magnetoresistance is several orders of magnitude smaller than the measured $\rho \; (T, B)$. When considering the multi-band behavior further fits of $\rho_{xy}$ using three-band model\cite{KimJS} reproduced $\rho_{xy}$ equally well. Moreover, nearly identical MR is obtained. Therefore, in either case multiband MR is much smaller than experiment results. This suggests possible contribution beyond the conventional orbital magnetoresistance that amplifies MR.

\begin{figure}
\centerline{\includegraphics[scale=0.3]{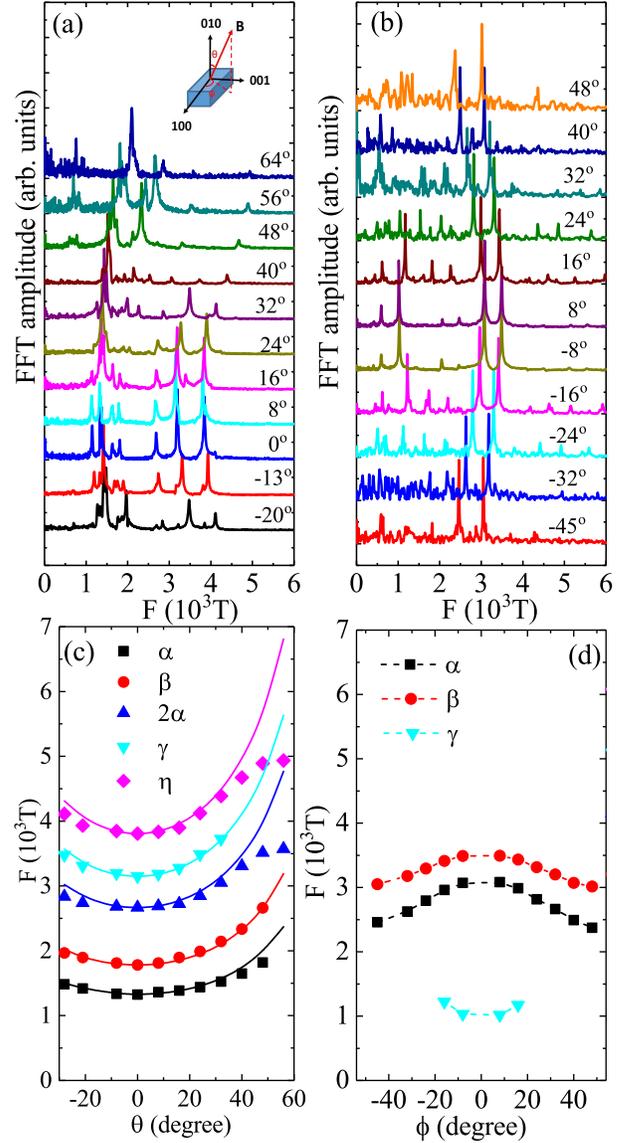}}
\caption{(Color online). FFT spectra of the SdH oscillation with field rotated in $bc$ plane (a) and FFT spectra of the dHvA with field rotated in the $ab$ plane (b). Angular dependence of the oscillation frequency for SdH (c), and dHvA (d). The spectra are nomalized and shifted vertically for clearity, solid lines are fitted with 2D model 1/cos($\theta$)}
\label{magnetism}
\end{figure}

MR for H//$b$, and cantilver for H//$a$ at different temperatures up to 18 T are shown in Fig. 4(a) and 4(b). MR shows non-saturating $H^n$ up to 18 T, consistent with result in Fig. 2. Clear oscillations were observed above 10 T. The oscillatory components are plotted as a function of 1/B in Fig. 4(c) and 4(d). Both Fig. 4(c) and 4(d) exhibit beat patterns, indicating that multiple frequencies contribute to oscillations. The fast Fourier transform (FFT) spectra of the oscillatory components are shown in Fig. 4(e) and 4(f). Four frequencies are observed in Fig. 4(e): 1384 T, 1799 T, 3183 T, and 3806 T. There are several peaks around $F_{\alpha}$ = 1384 T, suggesting contribution of several extrema. According the calculated Fermi surface of WP$_2$,\cite{Autes, Kumar} $F_{\alpha}$ = 1384 T, and $F_{\beta}$ = 1799 T can be ascribed to hole pockets, and $F_{\gamma}$ = 3183 T, and $F_{\eta}$ = 3806 T are from electron pockets.

From the Onsager relation, $F = (\Phi_0/2\pi^2)A_F$, where $\Phi_0$ is the flux quantum and $A_F$ is the orthogonal cross-sectional area of the Fermi surface, the Fermi surface is estimated to be 13 nm$^{-2}$, 17 nm$^{-2}$, 30 nm$^{-2}$, and 36 nm$^{-2}$, corresponding to 5\%, 7\%, 12\%, and 14\% of the total area of the Brillouin zone in the $ac$ plane.
Assuming the circular cross section of Fermi surface $A_F$ = $\pi$$k_F^2$, the band splitting induced by spin-orbital coupling can be estimated to be $k_{\gamma}$ - $k_{\eta}$ = 0.03 $\rm{\AA}$$^{-1}$, which is larger than that in MoP, and about half of that in giant Rashba effect material BiTeI.\cite{Shekhar, Ishizaka} Strong spin-orbital coupling in WP$_2$ indicates that spin texture, which could forbid backscattering, might play an important role in large magnetoresistance, similar to that in WTe$_2$.\cite{JiangJ}
For dHvA oscillation along $a$ axis, the Fermi surfaces from three frequencies 1217 T, 2949 T, and 3417 T are 12 nm$^{-2}$, 28 nm$^{-2}$, and 32 nm$^{-2}$, corresponding to 18\%, 42\%, and 48\% of the Brillouin zone.

In order to obtain the cyclotron mass for the main frequencies, the FFT amplitude $A$ versus temperature was fitted using Lifshitz-Kosevich formula,\cite{Shoeneberg} $A$$\sim$[$\alpha$$m^{*}$(T/B)/$\sinh$($\alpha$ $m^{*}$T/B)] where $\alpha$ = 2$\pi$$^2$$k_{\rm B}$/e$\hbar$ $\approx$ 14.69 T/K, $m^{*}$ = $m$/$m_{e}$ is the cyclotron mass ratio ($m_{e}$ is the mass of free electron). As shown in Fig. 4(e) and 4(f), the cyclotron mass along $b$ axis is close to the mass of free electron $m_e$, while cyclotron mass along $a$ axis is reduced to about 0.5$m_e$, which can be attributed to anisotropy of the Fermi surface. We estimate the Dingle temperatures from the FFT frequency with highest FFT peaks to be  2.4 K and 4.7 K for H//$b$ and H//$a$, respectively. The corresponding scattering times are 5.0 $\times$ 10$^{-13}$ s and 2.6 $\times$ 10$^{-13}$ s. Then, the mobility estimated by $\mu_q$ = $e\tau_q/m_c$ is 880 cm$^2$V$^{-1}$s$^{-1}$ and 458 cm$^2$V$^{-1}$s$^{-1}$, comparable to Fig. 3(c). The effective mass and mobility suggest that the Fermi surfaces detected by quantum oscillation are not under significant influence of the due the the Weyl points which are located below the Fermi level.\cite{Autes}

Angle dependence of SdH in $bc$ plane, and dHvA in $ab$ plane provides further insight into the shape of the Fermi surface. The FFT spectra of the SdH is presented in Fig. 5 (a), and dHvA in Fig. 5(b). For the SdH measurement, four main peaks are observed, the FFT peaks increases with the angle tilt from zero, and disappear above 60$^{\rm o}$. Solid lines in Fig.5(c) are fitted using 2D Fermi surface [$F(0)/cos(\theta)$]. Hence low angle quantum oscillations reveal quasi-2D character at low angles below 50$^{\rm o}$.
When the field is rotated in the $ab$ plane, with the field titled from $a$ axis, the frequency decreases with the angle, indicating the enlongated Fermi surface in the $bc$ plane.
The angle dependence of quantum oscillations agree well with the calculated Fermi surface.\cite{Autes, Kumar} The calculated Fermi surfaces of WP$_2$ consist of spaghetti-like hole Fermi surfaces and bow-tie-like electron Fermi surfaces, while the presence of strong spin-orbit coupling leads to splitting of Fermi surfaces.\cite{Autes, Kumar} The spaghetti-like hole Fermi surface extends along $b$ axis, bends along $a$ axis, while it is flat along $c$ axis, consistent with the quasi-2D enlonged $\alpha$ and $\beta$ bands in $bc$ plane. When magnetic field is applied in small angles around $b$ axis, SdH oscillations of $\gamma$ and $\eta$ bands are from the orbits across whole electron pockets in $ac$ plane, quasi-2D behavior of $\gamma$ and $\eta$ bands can be then attributed to relatively flat wall of electron pockets. Moreover, the appearance in pair of both the hole and electron Fermi surfaces with almost same angle dependence might be the result of band splitting effect which is induced by strong spin orbit coupling.

\section{CONCLUSIONS}

In conclusion, magnetotransport studies of WP$_{2}$ confirm the presence of multiple bands.\cite{Autes}  The large increase of resistivity in magnetic field cannot be explained by the multiband orbital magnetoresistance but the effective mass and mobility detected by quantum oscilllations are not under significant influence of Weyl points. Strong spin-orbit coupling effect was observed in dHvA whereas the temperature and angle dependence of dHvA measurements reveal anisotropic multiband characteristics and are in agreement with calculations. Even though crystal structure in WP$_2$ is not layered as in transition metal dichalcogenides, Fermi surfaces do show partial quasi-2D character.

\section*{Acknowledgements}

We thank John Warren for help with EDX measurements and Rongwei Hu for useful discussions. Work at BNL was supported by the U.S. DOE-BES, Division of Materials Science and Engineering, under Contract No. DE-SC0012704. Work at the National High Magnetic Field Laboratory is supported by the NSF Cooperative Agreement No. DMR-1157490, and by the state of Florida.

\emph{Note added.} During the review of this paper, a related study on WP$_2$ was reported by R. Sch\"{o}nemann \emph{et al}.,\cite{Schonemann} though the magnitude of magnetoresistance and RRR of the crystal is somewhat different, the Fermi surface and main conclusion is consistent with ours.


\begin{references}
\bibitem{Dirac} T.O. Wehling, A.M. Black-Schaffer and A.V. Balatsky, Advances in Physics \textbf{63}, 1 (2014).
\bibitem{WanX} Xiangang Wan, Ari M. Turner, Ashvin Vishwanath, and Sergey Y. Savrasov, Phys. Rev. B  \textbf{83}, 205101 (2011).
\bibitem{Balents} Leon Balents, Physics \textbf{4}, 36 (2011).
\bibitem{Soluyanov} A. A. Soluyanov, D. Gresch, Z. Wang, Q. Wu. M. Troyer, X. Dai, and B. A. Bernevig, Nature \textbf{527}, 495 (2015).
\bibitem{VafekO} Oskar Vafek and Ashvin Vishwanath, Annu. Rev. Condens. Matter Phys. \textbf{5}, 83 (2014).
\bibitem{BansilA} A. Bansil, Hsin Lin and Tanmoy Das, Rev. Mod. Phys. \textbf{88}, 021004 (2016).
\bibitem{HasanM}  M.Z. Hasan and C. L. Kane, Rev. Mod. Phys. \textbf{82}, 3045 (2010).
\bibitem{JiaS} Shuang Jia, Su-Yang Xu and M. Zahid Hasan, Nature Materials \textbf{15}, 1140 (2016).
\bibitem{BurkovA} A. A. Burkov, Nature Materials \textbf{15}, 1145 (2016).
\bibitem{LiuJ} Jianpeng Liu and David Vanderbilt, Phys. Rev. B \textbf{90} 155316 (2014).
\bibitem{ZhengH}Hao Zheng, Guang Bian, Guoqing Chang, Hong Lu, Su-Yang Xu, Guangqiang Wang, Tay-Rong Chang, Songtian Zhang, Ilya Belopolski, Nasser Alidoust, Daniel S. Sanchez, Fengqi Song, Horng-Tay Jeng, Nan Yao, Arun Bansil, Shuang Jia, Hsin Lin, and M. Z. Hasan, Phys. Rev. Lett. \textbf{117}, 266804 (2016).
\bibitem{DaiX} Xi Dai, Nature Materials \textbf{15}, 5 (2015).
\bibitem{WangKF} K. Wang, D. Graf, L. Li, L. Wang, and C. Petrovic, Sci. Rep. \textbf{4}, 7328 (2014).
\bibitem{YuanZ} Z. Yuan, H. Lu, Y. Liu, J. Wang, and S. Jia, Phys. Rev. B \textbf{93}, 184405 (2016).
\bibitem{Rundquist} S. Rundqvist and T. Lundstrom, Acta Chem. Scand. \textbf{17}, 37 (1963).
\bibitem{YoungS} S. M. Young and C. L. Kane, Phys. Rev. Lett. \textbf{115}, 126803 (2015).
\bibitem{BzdusekT} T. Bzdusek, QuaanSheng Wu, A. R\"{u}egg, M. Sigrist and A. A. Soluyanov, Nature \textbf{538}, 75 (2016).
\bibitem{ZhaoYX} Y. X. Zhao and A. P. Schnyder, Phys. Rev. B \textbf{94}, 195109 (2016).
\bibitem{Autes} G. Autes, D. Gresch, M. Troyer, A. A. Soluyanov, and O. V. Yazyev, Phys. Rev. Lett. \textbf{117}, 066402 (2016).
\bibitem{TaftiPNAS} F. F. Tafti, Q. Gibson, S. Kushwaha, J. W. Krizan, N. Haldolaarachchige, and R. J. Cava, Proc. Natl. Acad. Sci. USA \textbf{113}, E3475 (2016).
\bibitem{Ziman} M. Ziman, \textit{Electrons and Phonons} (Clarendon Press, Oxford, 1962).
\bibitem{Tafti} F. F. Tafti, Q. D. Gibson, S. K. Kushwaha, N. Haldolaarachchige, and R. J. Cava, Nature Phys. \textbf{12}, 272 (2015).
\bibitem{Hechang} Shanshan Sun, Qi Wang, Peng-Jie Guo, Kai Liu, and Hechang Lei, New J. Phys. \textbf{18}, 082002 (2016).
\bibitem{Cava} M. N. Ali, J. Xiong, S. Flynn, J. Tao, Q. D. Gibson, L. M. Schoop, T. Liang, N. Haldolaarachchige, M. Hirschberger, N. P. Ong, R. J. Cava, Nature \textbf{514}, 205 (2014).
\bibitem{ChenD} D. Chen, L. X. Zhao, J. B. He, H. Liang, S. Zhang, C. H. Li, L. Shan, S. C. Wang, Z. A. Ren, C. Ren, and G. F. Chen, Phys. Rev. B \textbf{94}, 174411 (2016).
\bibitem{HuangX} X. Huang, L. Zhao, Y. Long, P. Wang, D. Chen, Z. Yang, H. Liang, M. Xue, H. Weng, Z. Fang, X. Dai, and G. Chen, Phys. Rev. X \textbf{5}, 031023 (2015).
\bibitem{Hosen} M. M. Hosen, K. Dimitri, I. Belopolski, P. Maldonado, R. Sankar, N. Dhakal, G. Dhakal, T. Cole, P. M. Oppeneer, D. Kaczorowski, F. Chou, M. Z. Hasan, T. Durakiewicz, and M. Neupane, Phys. Rev. B \textbf{95}, 161101(R) (2017).
\bibitem{MunE} E. Mun, H. Ko, G. J. Miller, G. D. Samolyuk, S. L. Bud'ko, and P. C. Canfield, Phys. Rev. B \textbf{85}, 035135 (2012).
\bibitem{Soule} D. E. Soule, Phys. Rev. \textbf{112}, 698 (1958).
\bibitem{ZhuZ} Z. Zhu, A. Collaudin, B. Fauqu\'{e}, W. Kang, and K. Behnia, Nature Phys. \textbf{8}, 89 (2011).
\bibitem{BudkoS} S. L. Bud$'$ko, C. Petrovic, G. Lapertot, C. E. Cunningham,, P. C. Canfield, M.-H. Jung and A. H. Lacerda Phys. Rev. B \textbf{63}, 220503 (2001).
\bibitem{Cezairliyan} A. Cezairliyan, \textit{Specific Heat of Solids} Hemisphere Pub. Corp., New York (1988).
\bibitem{Khim} Seunghyun Khim, Klaus Koepernik, Dmitry V. Efremov, J. Klotz, T. F\"{o}rster, J. Wosnitza, Mihai I. Sturza, Sabine Wurmehl, Christian Hess, Jeroen van den Brink, and Bernd B\"{u}chner, Phys. Rev. B \textbf{94}, 165145 (2016).
\bibitem{Shoeneberg} D. Shoeneberg, \textit{Magnetic} $Oscillation$ $in$ $Metals$ (Cambridge University Press, Cambridge, 1984).
\bibitem{WangL} Lin Wang, Ignacio Gutierrez-Lezama, Celine Barreteau, Nicolas Ubrig, Enrico Giannini Alberto F. Morpurgo, Nature Comms. \textbf{6}, 8892 (2015).
\bibitem{NiuXH} X. H. Niu, D. F. Xu, Y. H. Bai, Q. Song, X. P. Shen, B. P. Xie, Z. Sun, Y. B. Huang, D. C. Peets, and D. L. Feng, Phys. Rev. B \textbf{94}, 165163 (2016).
\bibitem{WuY} Yun Wu, Tai Kong, Lin-Lin Wang, D. D. Johnson, Daixiang Mou, Lunan Huang, Benjamin Schrunk, S. L. Bud'ko, P. C. Canfield, and Adam Kaminski, Phys. Rev. B \textbf{94}, 081108 (2016).
\bibitem{NayakJ} Jayita Nayak, Shu-Chun Wu, Nitesh Kumar, Chandra Shekhar, Sanjay Singh, J\"{o}rg Fink, Emile E.D. Rienks, Gerhard H. Fecher, Stuart S.P. Parkin, Binghai Yan and Claudia Felser, Nature Comms. \textbf{8}, 13942 (2016).
\bibitem{WuY2} Yun Wu, Lin-Lin Wang, Eundeok Mun, D. D. Johnson, Daixiang Mou, Lunan Huang, Yongbin Lee, S. L. Bud$'$ko, P. C. Canfield and Adam Kaminski, Nature Physics \textbf{12}, 667 (2016).
\bibitem{MaoZQ} J. Hu, Z. Tang, J. Liu, X. Liu, Y. Zhu, D. Graf, K. Myhro, S. Tran, C. N. Lau, J. Wei, and Z. Mao, Phys. Rev. Lett. \textbf{117}, 016602 (2016).
\bibitem{Pippard} A. B. Pippard, $Magnetoresistance in metals$ (Cambridge University Press, Cambridge, 1989).
\bibitem{ZhangCL} C. L. Zhang, Z. Yuan, Q. D. Jiang, B. Tong, C. Zhang, X. C. Xie, and S. Jia, Phys. Rev. B \textbf{95}, 085202 (2017).
\bibitem{SmithR} R. A. Smith, \textit{Semiconductors} (Cambridge University Press, 1978).
\bibitem{KefengSr} Kefeng Wang, D. Graf, Hechang Lei, S. W. Tozer, and C. Petrovic, Phys. Rev. B \textbf{84}, 220401(R) (2011).
\bibitem{KefengCa} Kefeng Wang, D. Graf, Limin Wang, Hechang Lei, S. W. Tozer, and C. Petrovic, Phys. Rev. B \textbf{85}, 041101(R) (2012).
\bibitem{KrienerM} M. Kriener, A. Kikkawa, T. Suzuki, R. Akashi, R. Arita, Y. Tokura, and Y. Taguchi Phys. Rev. B \textbf{91}, 075205 (2015).
\bibitem{KimJS} J. S. Kim, J. Appl. Phys. \textbf{86}, 3187 (1999).
\bibitem{Kumar} N. Kumar, Y. Sun, K. Manna, V. S\"{u}ss, I. Leermakers, O. Young, T. F\"{o}rster, M. Schmidt, B. Yan, U. Zeitler, C. Felser, C. Shekhar, arXiv: 1703.04527
\bibitem{Shekhar} C. Shekhar, Y. Sun, N. Kumar, M. Nicklas, K. Manna,  V. S\"{u}$\beta$, O. Young, I. Leermakers, T. F\"{o}rster, M. Schmidt, L. Muechler, P. Werner, W. Schnelle, U. Zeitler, B. Yan, S. S. P. Parkin, C. Felser, arXiv: 1703.03736.
\bibitem{Ishizaka} K. Ishizaka, M. S. Bahramy, H. Murakawa, M. Sakano, T. Shimojima, T. Sonobe, K. Koizumi, S. Shin, H. Miyahara, A. Kimura, K. Miyamoto, T. Okuda, H. Namatame, M. Taniguchi, R. Arita, N. Nagaosa, K. Kobayashi, Y. Murakami, R. Kumai, Y. Kaneko, Y. Onose, and Y. Tokura, Nat. Mater. \textbf{10}, 521 (2011).
\bibitem{JiangJ} J. Jiang, F. Tang, X. C. Pan, H. M. Liu, X. H. Niu, Y. X. Wang, D. F. Xu, H. F. Yang, B. P. Xie, F. Q. Song, P. Dudin, T. K. Kim, M. Hoesch, P. K. Das, I. Vobornik, X. G. Wan, and D. L. Feng, Phys. Rev. Lett. \textbf{115}, 166601 (2015).
\bibitem{Schonemann} R. Sch\"{o}nemann, N. Aryal, Q. Zhou, Y. -C. Chiu, K. -W. Chen, T. J. Martin, G. T. McCandless, J. Y. Chan, E. Manousakis, and L. Balicas, arXiv: 1706.10135.





\end{references}
\end{document}